\documentclass[grl]{agutex}

\usepackage{amsfonts,amsmath,amsbsy}
\usepackage{graphicx}

\setkeys{Gin}{draft=false}

\def\pct{\%{ }}

\makeatletter
\renewcommand*\env@matrix[1][\arraystretch]{%
  \edef\arraystretch{#1}%
  \hskip -\arraycolsep
  \let\@ifnextchar\new@ifnextchar
  \array{*\c@MaxMatrixCols c}}
\makeatother

\authorrunninghead{Beron-Vera, Olascoaga \& Lumpkin}%
\titlerunninghead{Inertia-induced accumulation}%
\lefthead{Beron-Vera, Olascoaga \& Lumpkin}%
\righthead{Inertia-induced accumulation}%
\journalid{2016}%
\articleid{X}{X}%
\paperid{2016JGRXXXXX}%
\cpright{AGU}{2016}%
\received{\today}%
\revised{\today}%
\accepted{\today}%

\authoraddr{F.\ J.\ Beron-Vera, RSMAS/ATM, University
of Miami, 4600 Rickenbacker Cswy., Miami, FL 33149, USA.
(fberon@rsmas.miami.edu)}

\authoraddr{M.\ J.\ Olascoaga, RSMAS/OCE, University
of Miami, 4600 Rickenbacker Cswy., Miami, FL 33149, USA.
(jolascoaga@rsmas.miami.edu)}

\authoraddr{R.\ Lumpkin, NOAA/AOML, 4301 Rickenbacker
Cswy., Miami, FL 33149, USA.  (rick.lumpkin@noaa.gov)}

\begin{document}

\title{Inertia-induced accumulation of flotsam in the subtropical gyres}

\bigskip \authors{F.\ J.\ Beron-Vera\altaffilmark{1}, M.\ J.\
Olascoaga\altaffilmark{2}, R.\ Lumpkin\altaffilmark{3}}

\altaffiltext{1}{RSMAS/ATM, University of Miami.}
\altaffiltext{2}{RSMAS/OCE, University of Miami.}
\altaffiltext{3}{NOAA/AOML.}

\begin{abstract}
  Recent surveys of marine plastic debris density have revealed
  high levels in the center of the subtropical gyres.  Earlier
  studies have argued that the formation of great garbage patches
  is due to Ekman convergence in such regions.  In this work we
  report a tendency so far overlooked of drogued and undrogued
  drifters to accumulate distinctly over the subtropical gyres,
  with undrogued drifters accumulating in the same areas where
  plastic debris accumulate.  We show that the observed accumulation
  is too fast for Ekman convergence to explain it.  We demonstrate
  that the accumulation is controlled by finite-size and buoyancy
  (i.e., inertial) effects on undrogued drifter motion subjected
  to ocean current and wind drags. We infer that the motion of
  flotsam in general is constrained by similar effects.  This is
  done by using a newly proposed Maxey--Riley equation which models
  the submerged (surfaced) drifter portion as a sphere of the
  fractional volume that is submerged (surfaced).
\end{abstract}

\begin{article}

\section*{Key points}
\begin{itemize}
\item Undrogued drifters and plastic
  debris accumulate similarly in the subtropical gyres.
\item The accumulation is too fast to be due to Ekman convergence.
\item Inertial effects (i.e., of finite size and buoyancy) explain
the accumulation.
\end{itemize}  

\section{Introduction}  

The purpose of this brief communication is twofold.  First, we
report, for the first time, that drogued and undrogued drifters
tend to distribute differently in the subtropical gyres, with
undrogued drifters accumulating in regions where microplastic density
surveys indicate elevated levels of floating marine debris
\citep{Cozar-etal-14}.  Second, we provide an explanation for this
tendency using an appropriate reduced Maxey--Riley equation
\citep{Maxey-Riley-83, Cartwright-etal-10} for the motion of buoyant
finite-size (i.e., inertial) spherical particles.  Unlike the
standard Maxey--Riley equation, used previously in oceanographic
applications \citep{Tanga-Provenzale-94, Beron-etal-15}, the new
equation derived here takes into account the combined effects of
water and air drags. The water velocity is taken to be causally
related to the air velocity so the role of the Ekman transport in
the accumulation of flotsam in the ocean gyres, proposed earlier
\citep{Maximenko-etal-12}, can be unambiguously evaluated.  This
is attained by considering the water velocity as the surface ocean
velocity output from an ocean general circulation model.  The air
velocity is in turn obtained from the wind velocity that forces the
model.  The present approach is dynamical, aimed at explaining
observed behavior, and thus is fundamentally different than earlier
probabilistic approaches \citep{Maximenko-etal-12, vanSebille-etal-12},
more concerned with reproducing observations.

\section{Observed accumulation}
 
The drifter data belong to the NOAA (National Oceanic Atmospheric
Administration) Global Drifter Program over the period 1979--2015
\citep{Lumpkin-Pazos-07}.  The drifter positions are satellite-tracked
by the \emph{Argos} system or GPS (Global Positioning System). The
drifters follow the SVP (Surface Velocity Program) design, consisting
of a surface spherical float which is drogued at 15 m, to minimize
wind slippage and wave-induced drift \citep{Sybrandy-Niiler-91}.

The top-left panel of Fig.\ 1 shows density (expressed as number
per degree squared) of drifters after a period of at least 1 yr
from deployment for all drifters that remained drogued over the
entire period. The top-right panel shows the same but after at least
1 yr since the drifters lose their drogues.  The initial positions
(insets) are similarly homogeneously distributed.  But there is a
difference in the final positions:  the undrogued drifters reveal
a more clear tendency to accumulate in the subtropical gyres.  The
accumulation is most evident in the North and South Atlantic gyres
and the South Pacific gyre.

\begin{figure*}[t]
  \centering%
  \includegraphics[width=\textwidth]{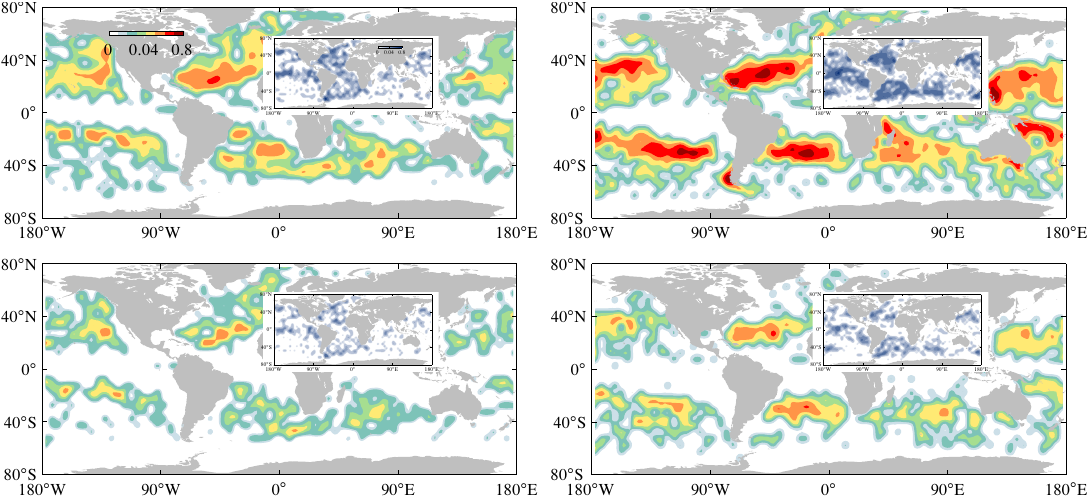}%
  \caption{Expressed as number per degree squared, density of drogued
  (left) and undrogued (right) drifters from the NOAA Global Drifter
  Program over 1979--2015 after at least 1 (top) or exactly 1.5
  (bottom) yr past the time at  deployment for drogued drifters or
  the location where a drifter loses the drogue (insets).}
\end{figure*}

The tendency of undrogued drifters to accumulate in the North and
South Atlantic gyres is particularly robust as it is not influenced
by the disparity in the amounts (1621 vs 2895) and mean lifetimes
(1.5 vs 2 yr) of the drogued and undrogued drifters used in the
construction of the top panels of Fig.\ 1.  This follows from the
inspection of the bottom panels, which show the same as in the top
panels but restricted to equal number of drifters (826) and length
of the trajectory records (1.5 yr).

The reported accumulation tendency had been inferred earlier, but
for both undrogued and drogued drifters and from the topology of
ensemble-mean streamlines constructed using drifter velocities
\citep{Maximenko-etal-12}.  These authors found their result
unexpected given  the different water-following characteristics of
drogued and undrogued drifters \citep{Niiler-Paduan-95}.  The
discrepancy with our finding may be attributed to errors in the
drogue presence verification, which were discovered at the time of
that publication and corrected in \citet{Lumpkin-etal-12}.

The accumulation inferred in \citet{Maximenko-etal-12} was explained
as a consequence of Ekman transport using steady flow arguments on
fluid particle motion.  In the next section we show that inertial
effects provide an explanation for the accumulation when spherical
float motion opposed by unsteady water and air flow drag is considered.

\section{Simulated accumulation}

Consider a small spherical particle of radius $a$ and density
$\rho_\mathrm{p} \le \rho$, where $\rho$ and $\rho_\mathrm{a}$ are
the water and air densities.  The fraction of water volume displaced
by the particle is $\delta^{-1}$, where
\begin{equation}
  \delta : = \frac{\rho}{\rho_\mathrm{p}}.
  \label{eq:delta}
\end{equation}

Posing the exact motion equation for a buoyant finite-size particle
immersed in a fluid in motion is a challenging task
\citep{Cartwright-etal-10}, which was solved to a very good
approximation by \citet{Maxey-Riley-83}.  As a first step toward
posing that for a particle at the air--sea interface, the more
complicated case of interest here, we proceed heuristically by
modeling the particle piece immersed in the water (air) as a sphere
of the fractional volume that is immersed in the water (air), and
assuming that it evolves according to the Maxey--Riley set.  Adding
the forces acting on each of the spheres as if they were decoupled
from one another, we obtain (cf.\ Supporting Information, Appendix
A):
\begin{equation}
  \ddot{x} + f\dot{x}^\perp = \mathrm{D}_t v + fv^\perp -
  \frac{2\left(\gamma + \sqrt[3]{\delta - 1}\right)}{3\gamma\sqrt[3]{\delta}\tau}
  \left(\dot{x} - u\right), \label{eq:mr}
\end{equation}
where
\begin{equation}
  u := \frac{\gamma v + \sqrt[3]{\delta - 1} v_\mathrm{a}}{\gamma + \sqrt[3]{\delta - 1}}.  
  \label{eq:u}
\end{equation}
Here $x$ denotes position on the horizontal plane; $v$ and
$v_\mathrm{a}$ are water and air velocities; $\mathrm{D}_t :=
\partial_t + v \cdot \nabla$; $f$ is the Coriolis parameter; and
\begin{equation}
  \tau := \frac{2a^2}{9\nu\delta},\quad 
  \gamma := \frac{\nu\rho}{\nu_{\mathrm{a}}\rho_\mathrm{a}},
  \label{eq:taugamma}
\end{equation}
where $\nu$ and $\nu_\mathrm{a}$ are water and air dynamic viscosities.
The left-hand-side of (\ref{eq:mr}) is the absolute acceleration
of the particle.  The first and second terms on the right-hand-side
are flow and drag forces mediated by added mass effects, respectively,
which water and air exert on the particle.

The Maxey--Riley equation (\ref{eq:mr}) constitutes a nonautonomous
four-dimensional dynamical system for the particle position and
velocity, $v_\mathrm{p} = \dot{x}$.  To integrate it one must specify
both initial particle position and velocity, which is not known in
general.  In addition,  long reversed-time integrations of
(\ref{eq:mr}), which are useful for instance in pollution source
detection, are not feasible because  the term $u/\tau$ tends to
cause numerical instability as it has been noted earlier for the
standard Maxey--Riley equation \citep{Haller-Sapsis-08}.

However, for a sufficiently small particle ($\tau \to 0$) the
Maxey--Riley equation (\ref{eq:mr}) reduces to (cf.\ Supporting
Information, Appendix B)
\begin{equation}
  \dot{x} = v_\mathrm{p} = u +
  \frac{3\gamma\sqrt[3]{\delta}\tau}{2\left(\gamma
  + \sqrt[3]{\delta - 1}\right)}
  \left(\mathrm{D}_t v + fv^\perp - fu^\perp\right).  
  \label{eq:rmr}
\end{equation}
This equation, which will be referred to as an \emph{inertial
equation}, constitutes a two-dimensional system in the position,
and thus can be integrated without knowledge of the initial velocity.
Also, it is not subjected to numerical instability in long backward-time
integration.  Up to an $O(\tau^2)$ error, the particle velocity is
equal to a weighted average of the water and air velocities ($u$)
plus a term proportional to the water absolute and Coriolis (due
to $u$) accelerations times $\tau$, the particle timescale.

To carry out the integration of (\ref{eq:rmr}) realizations of $v$
and $v_\mathrm{a}$ near the ocean--atmosphere interface are needed.
Here we have chosen to consider $v$ as given by surface ocean
velocity from the Global $1/12^\circ$ HYCOM (HYbrid-Coordinate Ocean
Model) $+$ NCODA (Navy Coupled Ocean Data Assimilation) Ocean
Reanalysis (GLBu0.08/expt{\textunderscore}19.0)
\citep{Cummings-Smedstad-13}.  In turn, we take $v_\mathrm{a}$ as
the wind velocity from the National Centers for Environmental
Prediction (NCEP) Climate Forecast System Reanalysis (CFSR), which
is employed to construct the wind stress applied on the model.

Also needed for the integration of (\ref{eq:rmr}) are estimates of
parameters $\gamma$, $\delta$, and $\tau$.  For typical water and
air density and viscosity values, $\gamma \approx 60$. From the
configuration of the undrogued drifters we infer $\delta = 2$ (about
half of the spherical float is submerged when the drogue is not
present) and further estimate $\tau \approx 0.05$ d (the mean radius
of the float is about 17.5 cm). Note that $\tau$ is small compared
to relevant timescales such as the turnover time of a mesoscale
eddy (a few days) or a subtropical gyre (a few years) \citep{Vallis-06}.

For the above velocity realizations and parameter choices we begin
by integrating the inertial equation (\ref{eq:rmr}) using a
Runge--Kutta method from a uniform distribution of particles. For
comparison we also integrate $\dot{x} = v$ from the same positions.
In both cases integrations are initialized along three simulation
years (2005--2007).  Starting on the drifter deployment positions
or where the drifters lose their drogues on the corresponding dates
is not possible because model output is not available over the
entire drifter trajectory records.  The proposed ensemble integrations
facilitate intercomparisons and also guarantee robustness of the
results. The integrations are carried over a period of 4 yr, which
is sufficiently long  to reveal accumulation (undrogued drifters
show clear signs of accumulation after about 1.5 yr).  For convenience
we restrict the analysis to the North Atlantic; similar results are
attained in the other basins.

The density (normalized by initial density) of inertial particles,
i.e., controlled by (\ref{eq:rmr}), after 1.5 (Fig.\ 2, top-left
panel) and 4 (Fig.\ 2, middle-left panel) yr is high in the center
of the subtropical gyre  as is that of undrogued drifters . Note
also (in the insets) that particles controlled by the Maxey--Riley
equation (\ref{eq:mr}) (circles) take  similar final distributions
as inertial particles (dots).  This confirms the validity of
(\ref{eq:rmr}), which attracts solutions of (\ref{eq:mr}).

\begin{figure*}[t]
  \centering
  \includegraphics[width=\textwidth]{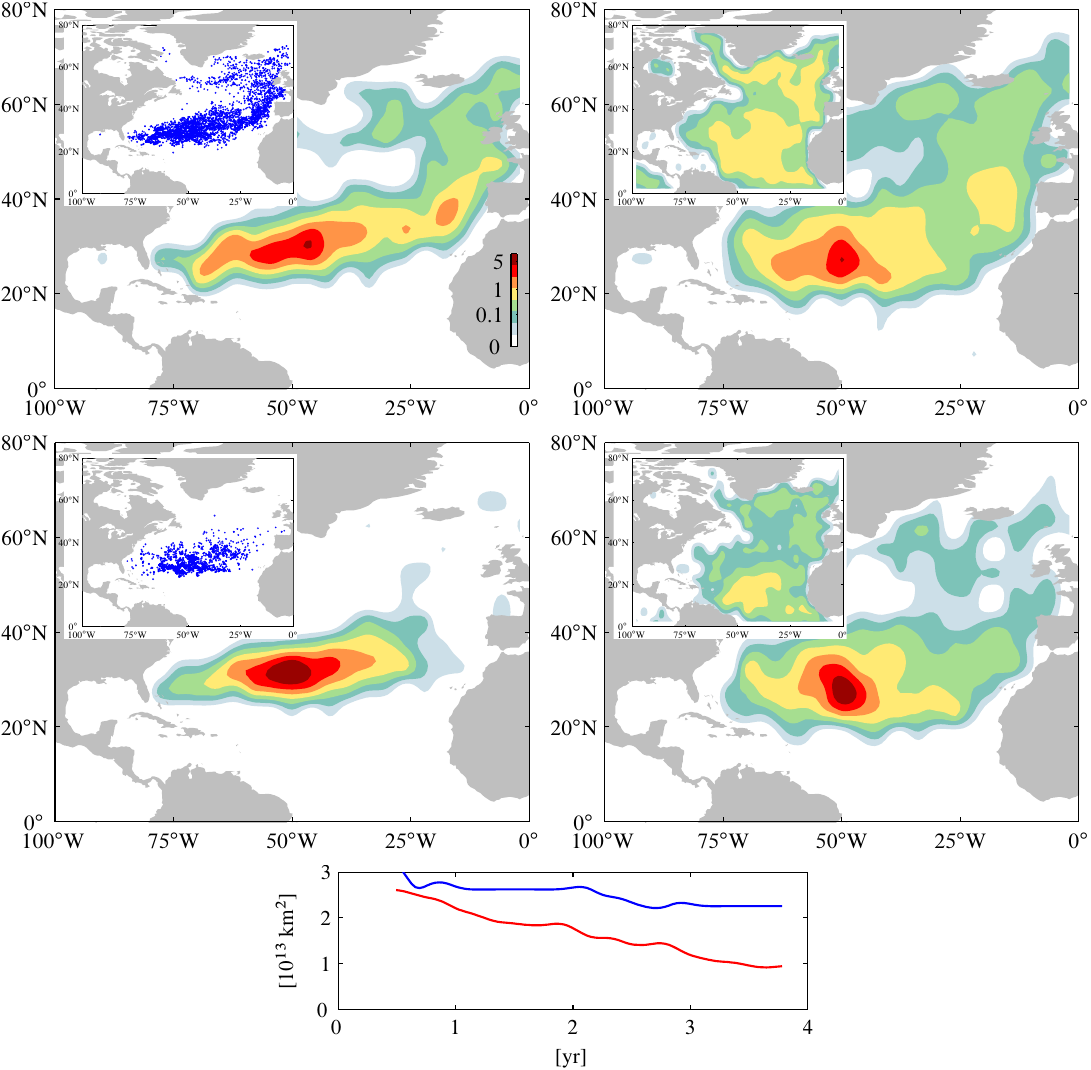}%
  \caption{Density of particles after 1.5 (top) and 4 (middle) yr
  of integration of the inertial equation (\ref{eq:rmr}) (left) and
  of advection by water velocity (right) normalized by density in
  the initially uniform distribution of particles.  Insets in the
  left show final positions of inertial particles (dots) and particles
  obeying the full Maxey--Riley equation (\ref{eq:mr}) (circles).
  Insets in the right show normalized density for particles advected
  by velocity derived geostrophically from sea-surface height.  In
  the bottom panel, as function of time, area of the region where
  normalized particle density is higher than 1\pct for inertial
  (red) and water (blue) particles.  Water velocity is given by
  surface ocean velocity from the 1/12$^\circ$ Global HYCOM$+$NCOM
  Ocean Reanalysis, from which sea-surface height is also taken.
  The air velocity corresponds to the wind velocity from the NCEP/CFSR
  reanalysis used to construct the wind stress that forces the
  model.}
\end{figure*} 

By contrast, after 1.5 yr water particles, i.e., controlled by
$\dot{x} = v$, take a more homogeneous distribution (Fig.\ 2,
top-right panel), which is in better agreement with the distribution
taken by drogued drifters .  Accumulation in this case, most evident
after 4 yr (Fig.\ 2, middle-right panel), can be attributed to Ekman
transport by comparing these distributions with the much more
homogeneous distributions attained by the particles when $v$ is
taken as geostrophic (i.e., divergenceless) velocity inferred from
the model sea surface height (SSH) field (insets).

Accumulation due to Ekman transport is a slow process.  This is
evident from the inspection of the bottom panel of Fig.\ 2, which
shows that the region where normalized particle density is higher
than 1\pct decays nearly two times faster for inertial particles
than for water particles. While there are not enough sufficiently
long drogued drifter trajectories to verify this behavior, application
of a probabilistic approach similar to that used earlier
\citep{Maximenko-etal-12, vanSebille-etal-12} on such drifters
suggests it .
  
The behavior of the inertial particles just described can be
anticipated by considering an idealized model of the large-scale
circulation in the North Atlantic.  In the simplest such models,
due to \citet{Stommel-66}, the slow steady flow is divergenceless
($\nabla\cdot v = 0$) and has an anticyclonic basin-wide gyre,
driven by strong steady
 westerlies and trade winds, so
$\nabla\cdot v_\mathrm{a} = 0$.  Under such conditions,
\begin{equation}
  \nabla \cdot v_\mathrm{p} \approx
  \frac{3\gamma\sqrt[3]{\delta}\tau}{2\left(\gamma
  + \sqrt[3]{\delta - 1}\right)^2}
  \sqrt[3]{\delta - 1} f\omega_\mathrm{a}, 
  \label{eq:div}
\end{equation}
where $\omega_\mathrm{a} := - \nabla\cdot v_\mathrm{a}^\perp$ is
the air vorticity.  Because $f\omega_\mathrm{a} < 0$, (\ref{eq:div})
is negative, which promotes accumulation of inertial particles in
the center of the gyre.  In other words, inertia-induced accumulation
occurs on a faster timescale than Ekam convergence, which is a
higher order (in the Rossby number) effect in Stommel's model.

A pertinent question is if undrogued drifter accumulation may be
inferred by simply considering $\dot{x} = v + \alpha\,v_\mathrm{a}$
with $\alpha > 0$ small, an ad-hoc model widely used to simulate
windage effects on floating matter in the ocean \citep{Duhec-etal-15}.
Use of $\alpha = 0.01$ suggests that it may indeed be possible after
1.5 yr of evolution (Fig.\ 3, left panel), but use of a slightly
larger value within the commonly used range such as $\alpha = 0.05$
reveals leakage of particles in the southwest direction (Fig.\ 3,
middle panel).  This emphasizes the importance of finite-size
effects.  More specifically, for undrogued drifter parameters $u
\approx 0.99\,v + 0.02\,v_\mathrm{a}$ in (\ref{eq:u}), which
incidentally is close to the ad-hoc models just considered.  This
is the first term in the inertial equation (\ref{eq:rmr}).  Finite-size
effects are accounted for in the second term.

\begin{figure*}[t]
  \centering
  \includegraphics[width=\textwidth]{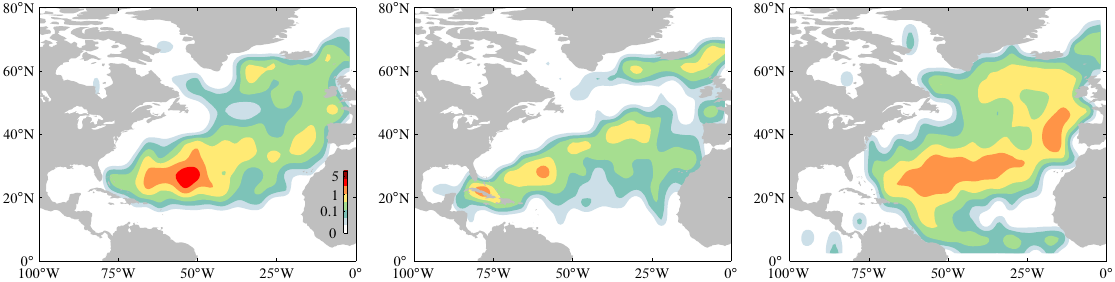}%
  \caption{As in in the top panels of Fig.\ 2, but for particles
  advected for 1.5 yr using model velocity with 1\pct (right) and
  5\pct (middle) windage added, and obeying the inertial equation
  (\ref{eq:rmr}) with the water velocity derived geostrophically
  from the model sea-surface height output (right).}
\end{figure*}

Another relevant question is if accumulation is revealed when
SSH-derived instead of full $v$ is used in the inertial equation
(\ref{eq:rmr}).  The relevance of this question stems from the wide
use of satellite altimetry measurements of SSH in diagnosing surface
ocean currents \citep{Fu-etal-10}.  The answer to the question is
partially affirmative as the final particle position distribution
reveals (Fig.\ 3, right panel).  A similar result is attained if
altimetric SSH is employed.

\section{Concluding remarks}

We conclude that undrogued drifters are  quite strongly influenced
by inertial effects.  Drogued drifters, by contrast, appear to more
closely follow the water motion.  Ekman transport contributes to
accumulate water, but this acts on a longer time scale than inertial
effects.  We infer, then, that marine plastic debris, which accumulate
in the same places as undrogued drifters, and flotsam in general
must be affected by inertial effects in a similar manner as undrogued
drifters.

We close by noting that shipwreck and airplane debris tracking,
pollution source identification, and search and rescue operations
at sea are among the many practical applications that may benefit
from the use of the inertial equation derived here.  Whether
inaccuracies resulting from our heuristic derivation of this equation
and omission of a number of potentially important processes (Stokes
drift, infragravity waves, subgrid motions, etc.) or the quality
of the velocity realizations will constrain more its success  in
such applications is a subject of ongoing research.

\begin{acknowledgments}
   We thank the comments by an anonymous reviewer, which have helped
   us to clarify the derivation of the inertial equation. The drifter
   data were collected by the NOAA Global Drifter Program
   (http://\allowbreak www.\allowbreak aoml.\allowbreak noaa.\allowbreak
   gov/\allowbreak phod/dac). The $1/12^\circ$ Global HYCOM$+$NCODA
   Ocean Reanalysis was funded by the U.S.\ Navy and the Modeling
   and Simulation Coordination Office. Computer time was made
   available by the DoD High Performance Computing Modernization
   Program. The output and forcing are publicly available at
   http://hycom.org. Our work was supported by CIMAS and the Gulf
   of Mexico Research Initiative (FJBV and MJO), and NOAA/AOML (RL).
\end{acknowledgments}

\appendix

\section{The Maxey--Riley equation}

The exact motion of  inertial particles in the flow of a fluid is
controlled by the Navier--Stokes equation with moving boundaries
as such particles are extended objects in the fluid with their own
boundaries [\emph{Cartwright et al.}, 2010]. This approach results
in complicated partial differential equations which are very
difficult---if not impossible---to solve and interpret. However, a
good approximation to the motion of a small spherical particle,
formulated in terms of an ordinary differential equation (ODE), is
provided by the Maxey--Riley equation [\emph{Maxey and Riley},
1983].  Assuming that the time it takes such a particle to revisit
a given region is long, so the Basset history term can be ignored,
such equation is a classical mechanics Newton equation of the form:
\begin{equation}
  m_\mathrm{p} a_\mathrm{p}  = \mathcal{F}_\mathrm{FF} +
  \mathcal{F}_\mathrm{AM} + \mathcal{F}_\mathrm{SD}.  
  \label{eq:newton}
\end{equation}

On the left-hand-side of \eqref{eq:newton}, $m_\mathrm{p} = \frac{4}{3}
\pi a^3 \rho_\mathrm{p}$ is the mass of the particle and $a_\mathrm{p}$
is its acceleration.  In the case of interest where the fluid is
geophysical such acceleration is the absolute acceleration, which
is given by [\emph{Provenzale}, 1999; \emph{Beron-Vera et al.},
2015]:
\begin{equation}
  a_\mathrm{p} = \ddot{x} + f \dot{x}^\perp,
\end{equation}
where $x$ denotes position on a plane tangent to the Earth, which
rotates with angular velocity $\frac{1}{2}f$. (The centrifugal
acceleration was also included in \emph{Provenzale} [1999] , but
this is actually balanced out by the gravitational acceleration on
the plane.  Also, geometric terms due to the Earth sphericity are
omitted for simplicity.  While trajectory details depend on such
terms [\emph{Ripa}, 1997], they do not affect the results of the
paper. For completeness, the full spherical forms of Maxey--Riley
and inertial equations are given in Appendix C.

On the right-hand-side of \eqref{eq:newton}, $\mathcal{F}_\mathrm{SD}$
denotes the drag force.  Assuming that the flow is laminar, this
force is described by the Stokes law.  Deriving an exact expression
for a spherical drifter at the water--air interface will require
one to write the projected area of the submerged (surfaced) part
in terms of the water-to-particle density ratio, $\delta$,  whose
inverse determines the submerged water volume, but this does not
seem feasible or simple at least.  So we approach the problem
heuristically by modelling the submerged portion of the drifter as
a sphere of the fractional volume that is submerged and the surfaced
piece as another sphere of the fractional volume that is surfaced.
The radius of the sphere immersed in the water is
$\sqrt[3]{\delta^{-1}}\,a$, while that of the sphere immersed in
the air is $\sqrt[3]{1-\delta^{-1}}\,a$.  Neglecting Faxen corrections,
which is reasonable for a small particle, we write the drag force
as a superposition of the drag forces acting on the spheres as if
they were well separated from one another:
\begin{equation}
  \mathcal{F}_\mathrm{SD} = \sqrt[3]{\delta^{-1}}6\pi a\nu\rho(v -
  \dot{x}) + \sqrt[3]{1 - \delta^{-1}} 6\pi
  a\nu_\mathrm{a}\rho_\mathrm{a}(v_\mathrm{a} - \dot{x}).
  \label{eq:drag}
\end{equation}
Force $\mathcal{F}_\mathrm{AM}$ describes added mass effects, which
in the present case are due to the displacement of both water and
air as the particle moves.  To write an exact expression for this
force one should know the potential flow around a sphere at the
interface between two fluids in motion, which is not known in closed
form.  Thus we proceed as above and write it as a superposition of
the added masses by the two spheres.  This leads to the
following explicit form:
\begin{align}
  \mathcal{F}_\mathrm{AM} = {} & - \frac{1}{2}\delta^{-1} m \left(
  \ddot{x} + f\dot{x}^\perp - \mathrm{D}_t v - f v^\perp\right)
  \nonumber\\ & - \frac{1}{2}(1 - \delta^{-1}) m_\mathrm{a} \left(
  \ddot{x} + f\dot{x}^\perp - \mathrm{D}_{\mathrm{a}t} v_\mathrm{a}
  - f v_\mathrm{a}^\perp\right),
\end{align}
where $\mathrm{D}_{\mathrm{a}t} := \partial_t + v_\mathrm{a} \cdot
\nabla$, $m = \frac{4}{3}\pi a^3\rho$, and $m_\mathrm{a} =
\frac{4}{3}\pi a^3\rho_\mathrm{a}$.  Finally, term
$\mathcal{F}_\mathrm{FF}$ is the flow force, here exerted by both
water and air on the particle.  Proceeding as before, this reads:
\begin{equation}
  \mathcal{F}_\mathrm{FF} = \delta^{-1} m \left(\mathrm{D}_t
  v + fv^\perp\right) + (1- \delta^{-1}) m_\mathrm{a}
  \left(\mathrm{D}_{\mathrm{a}t} v_\mathrm{a} +
  fv_\mathrm{a}^\perp\right).
\end{equation}

With the above explicit forms of each of the terms in \eqref{eq:newton},
and taking into account that $\rho \gg \rho_\mathrm{a}$, the
Maxey--Riley equation (2) follows.

\section{The inertial equation}

The second-order ODE (2) is equivalent to the following
first-order ODE set:
\begin{equation}
  \dot{x} = v_\mathrm{p},\quad \dot{v}_\mathrm{p} = \mathrm{D}_t v
  + f(v - v_\mathrm{p})^\perp - \frac{2\left(\gamma +
  \sqrt[3]{\delta - 1}\right)}{3\gamma\sqrt[3]{\delta}\tau}(v_\mathrm{p} - u).
  \label{eq:mr2}
\end{equation}
When $\tau = 0$, $v_\mathrm{p} = u$, which suggests the asymptotic
series expansion $v_\mathrm{p} = u + u_1 + u_2 + \cdots$ where $u_n
= O(\tau^n)$. Plugging this series into the right-hand-side
equation of system \eqref{eq:mr2} and equating $O(\tau)$ terms, it
follows that
\begin{equation}
  u_1 = \frac{3\gamma\sqrt[3]{\delta}}{2\left(\gamma + \sqrt[3]{\delta
  - 1}\right)} \left(\mathrm{D}_t v + fv^\perp - fu^\perp\right)
  \label{eq:u1}.
\end{equation}
Inserting this expression in the left-hand-side equation of system
\eqref{eq:mr2}, the inertial equation (5), valid up to an $O(\tau^2)$
error, follows.  For the interpretation of (5) as a slow manifold
of \eqref{eq:mr2}, cf.\ \emph{Haller and Sapsis} [2008].

A few remarks relating to limiting behavior of the inertial equation
(5) are in order.  A sizeless ($\tau = 0$) neutrally-buoyant ($\delta
= 1$) particle behaves as expected as a fluid particle because (2)
reduces in this limit to
\begin{equation}
  \dot{x} = v_\mathrm{p} = v.
\end{equation}
Sizeless ($\tau = 0$) but buoyant ($\delta > 1$) particles obey
\begin{equation}
  \dot{x} = v_\mathrm{p} = u,
\end{equation}
where $u$ a weighted average of the air and water velocities.  When
a particle is completely exposed to the air above the water ($\delta
\to \infty$), the inertial equation (5) reduces to
\begin{equation}
  \dot{x} = v_\mathrm{p} = v_\mathrm{a},
\end{equation}
i.e., its motion is completely driven by the air velocity as expected.
A neutrally buoyant particle ($\delta = 1$) obeys 
\begin{equation}
  \dot{x} = v_\mathrm{p} = v + \frac{3}{2}\tau \mathrm{D}_t v,
\end{equation}
i.e., its behavior differs from that of a fluid particle unless
$\mathrm{D}_t v$ can be neglected in front of $v$ (the water flow
is nearly geostrophic).  If the air above the water is replaced by
vacuum ($\gamma \to \infty$), in which case particle motion is
opposed by water drag exclusively,
\begin{equation}
  \dot{x} = v_\mathrm{p} = v + \frac{3}{2}\sqrt[3]{\delta}\tau
  \mathrm{D}_t v.
\end{equation}
When the air is quiescent ($v_\mathrm{a} = 0$),
\begin{align}
  \dot{x} = {} &  v_\mathrm{p} = \frac{\gamma v}{\gamma + \sqrt[3]{\delta
  - 1}} \nonumber\\ 
  & + \frac{3\gamma\sqrt[3]{\delta}\tau}{2\left(\gamma +
  \sqrt[3]{\delta - 1}\right)} \left(\mathrm{D}_t v + \frac{\sqrt[3]{\delta
  - 1}}{\gamma + \sqrt[3]{\delta - 1}}fv^\perp \right).  
  \label{eq:va0}
\end{align}
This is different than the vacuum case as particle motion is opposed
by both water and air drags.

Finally, the inertial equation derived by \emph{Beron-Vera et al.}
[2015] does not follow as a limiting case of the inertial equation
derived here as that one implicitly assumes that the particle is
completely immersed in the water (density stratification must be
allowed in order for nearly horizontal motion to be possible in
such a case).  However, the attracting (repelling) role of cyclonic
(anticyclonic) coherent Lagrangian eddies [\emph{Haller and
Beron-Vera}, 2013; 2014] which requires the water flow to be nearly
geostrophic, predicted for floating objects is still realized for
sufficiently calm air.  In a such case $\mathrm{D}_t v$ can be
neglected in right-hand-side of \eqref{eq:va0}, so
\begin{equation}
  \nabla \cdot v_\mathrm{p} = - \frac{3\gamma\sqrt[3]{\delta(\delta
  - 1)}\tau}{2\left(\gamma + \sqrt[3]{\delta - 1}\right)^2} f\omega,
  \label{eq:div2}
\end{equation}
where $\omega = - \nabla \cdot v^\perp$ is the water vorticity.
Note that the sign of \eqref{eq:div2} is determined by $f\omega$,
and thus the conclusions that follow from (11) in \emph{Beron-Vera
et al.} [2015] for light particles (i.e., with $\delta > 1$) follow
from \eqref{eq:div2} as well.

\section{The equations in full spherical geometry}\label{app:geo}

Let $\lambda$ ($\vartheta$) be longitude (latitude) and $R$ the
mean Earth radius.  Then the Maxey--Riley equation reads:
\begin{align}
  \dot{\lambda}   = {} & R^{-1}\sec\vartheta\,v_\mathrm{p}^\lambda,\\
  \dot{\vartheta} = {} & R^{-1}v_\mathrm{p}^\vartheta\\
  \dot{v}_\mathrm{p}^\lambda = {} & \mathrm{D}_t v^\lambda + f
  (v_\mathrm{p}^\vartheta - v^\vartheta) + \psi (v_\mathrm{p}^\lambda
  v_\mathrm{p}^\vartheta - v^\lambda v^\vartheta)\nonumber\\ & -
  \frac{2(\gamma + \sqrt[3]{\delta - 1})}{3\gamma\sqrt[3]{\delta}\tau}
  \big(v_\mathrm{p}^\lambda - u^\lambda\big),\\
  \dot{v}_\mathrm{p}^\vartheta = {} & \mathrm{D}_t v^\vartheta + f
  (v^\lambda - v_\mathrm{p}^\lambda) + \psi(v^\lambda v^\lambda -
  v_\mathrm{p}^\lambda v_\mathrm{p}^\lambda)\nonumber\\ & -
  \frac{2(\gamma + \sqrt[3]{\delta - 1})}{3\gamma\sqrt[3]{\delta}\tau}
  \big(v_\mathrm{p}^\vartheta - u^\vartheta\big),
\end{align}
where  $\psi : =  R^{-1}\tan\vartheta$ and $\mathrm{D}_{t} =
\partial_t + R^{-1}\sec\vartheta \,v^\lambda\partial_\lambda +
R^{-1}v^\vartheta\partial_\vartheta$.

 The inertial equation, in turn, takes the form:
\begin{align}
  \dot{\lambda} = {} & R^{-1}\sec\vartheta\bigg(u^\lambda +
  \frac{3\gamma\sqrt[3]{\delta}\tau}{2(\gamma + \sqrt[3]{\delta -
  1})} \Big(\mathrm{D}_t v^\lambda\nonumber\\ & + f (u^\vartheta -
  v^\vartheta) + \psi(u^\lambda u^\vartheta - v^\lambda
  v^\vartheta)\Big)\bigg),\\ \dot{\vartheta} = {} & R^{-1}\bigg(u^\vartheta
  + \frac{3\gamma\sqrt[3]{\delta}\tau}{2(\gamma +\sqrt[3]{ \delta
  - 1})} \big(\mathrm{D}_t v^\vartheta\nonumber\\ & + f (v^\lambda
  - u^\lambda) + \psi(v^\lambda v^\lambda - u^\lambda
  u^\vartheta)\Big)\bigg).
\end{align}

\bibliographystyle{agu08}

\begin{thebibliography}{21}
\providecommand{\natexlab}[1]{#1}
\expandafter\ifx\csname urlstyle\endcsname\relax
  \providecommand{\doi}[1]{doi:\discretionary{}{}{}#1}\else
  \providecommand{\doi}{doi:\discretionary{}{}{}\begingroup
  \urlstyle{rm}\Url}\fi

\bibitem[{\textit{Beron-Vera et~al.}(2015)\textit{Beron-Vera, Olascoaga,
  Haller, Farazmand, {Tri\~nanes}, and Wang}}]{Beron-etal-15}
Beron-Vera, F.~J., M.~J. Olascoaga, G.~Haller, M.~Farazmand, J.~{Tri\~nanes},
  and Y.~Wang (2015), {Dissipative inertial transport patterns near coherent
  Lagrangian eddies in the ocean}, \textit{Chaos}, \textit{25}, 087,412,
  \doi{10.1063/1.4928693}.

\bibitem[{\textit{Cartwright et~al.}(2010)\textit{Cartwright, Feudel,
  K\'arolyi, {de Moura}, Piro, and T\'el}}]{Cartwright-etal-10}
Cartwright, J.~H.~E., U.~Feudel, G.~K\'arolyi, A.~{de Moura}, O.~Piro, and
  T.~T\'el (2010), Dynamics of finite-size particles in chaotic fluid flows, in
  \textit{Nonlinear Dynamics and Chaos: Advances and Perspectives}, edited by
  {M. Thiel et al.}, pp. 51--87, Springer-Verlag Berlin Heidelberg.

\bibitem[{\textit{Cozar et~al.}(2014)}]{Cozar-etal-14}
Cozar, A., et~al. (2014), Plastic debris in the open ocean, \textit{Proc. Nat.
  Acad. Sci. USA}, \textit{111}(28), 10,239--10,244,
  \doi{10.1073/pnas.1314705111}.

\bibitem[{\textit{Cummings and Smedstad}(2013)}]{Cummings-Smedstad-13}
Cummings, J.~A., and O.~M. Smedstad (2013), Variational data analysis for the
  global ocean, in \textit{Data Assimilation for Atmospheric, Oceanic and
  Hydrologic Applications}, vol.~2, edited by S.~K. Park and L.~Xu, chap.~13,
  Springer-Verlag Berlin Heidelberg, \doi{10.1007/978-3-642-35088-7-13}.

\bibitem[{\textit{Duhec et~al.}(2015)\textit{Duhec, Jeanne, Maximenko, and
  Hafner}}]{Duhec-etal-15}
Duhec, A.~V., R.~F. Jeanne, N.~Maximenko, and J.~Hafner (2015), {Composition
  and potential origin of marine debris stranded in the Western Indian Ocean on
  remote Alphonse Island, Seychelles}, \textit{Mar. Poll. Bull.},
  \textit{96}(1--2), 76--86, \doi{10.1016/j.marpolbul.2015.05.042}.

\bibitem[{\textit{Fu et~al.}(2010)\textit{Fu, Chelton, {Le Traon}, and
  Morrow}}]{Fu-etal-10}
Fu, L.~L., D.~B. Chelton, P.-Y. {Le Traon}, and R.~Morrow (2010), Eddy dynamics
  from satellite altimetry, \textit{Oceanography}, \textit{23}, 14--25.

\bibitem[{\textit{Haller and Beron-Vera}(2013)}]{Haller-Beron-13}
Haller, G., and F.~J. Beron-Vera (2013), {Coherent Lagrangian vortices: The
  black holes of turbulence}, \textit{J. Fluid Mech.}, \textit{731}, R4,
  \doi{10.1017/jfm.2013.391}.

\bibitem[{\textit{Haller and Beron-Vera}(2014)}]{Haller-Beron-14}
Haller, G., and F.~J. Beron-Vera (2014), {Addendum to `Coherent Lagrangian
  vortices: The black holes of turbulence'}, \textit{J. Fluid Mech.},
  \textit{755}, R3.

\bibitem[{\textit{Haller and Sapsis}(2008)}]{Haller-Sapsis-08}
Haller, G., and T.~Sapsis (2008), Where do inertial particles go in fluid
  flows?, \textit{Physica D}, \textit{237}, 573--583.

\bibitem[{\textit{Lumpkin and Pazos}(2007)}]{Lumpkin-Pazos-07}
Lumpkin, R., and M.~Pazos (2007), {Measuring surface currents sith Surface
  Velocity Program driftres: the instrument, its data and some recent results},
  in \textit{Lagrangian Analysis and Prediction of Coastal and Ocean Dynamics},
  edited by A.~Griffa, A.~D. Kirwan, A.~Mariano, T.~\"Ozg\"okmen, and
  T.~Rossby, chap.~2, pp. 39--67, Cambridge University Press.

\bibitem[{\textit{{Lumpkin} et~al.}(2012)\textit{{Lumpkin}, {Grodsky},
  {Centurioni}, {Rio}, {Carton}, and {Lee}}}]{Lumpkin-etal-12}
{Lumpkin}, R., S.~A. {Grodsky}, L.~{Centurioni}, M.-H. {Rio}, J.~A. {Carton},
  and D.~{Lee} (2012), Removing spurious low-frequency variability in drifter
  velocities, \textit{J. Atm. Oce. Tech.}, \textit{30}, 353--360,
  \doi{10.1175/JTECH-D-12-00139.1}.

\bibitem[{\textit{Maxey and Riley}(1983)}]{Maxey-Riley-83}
Maxey, M.~R., and J.~J. Riley (1983), Equation of motion for a small rigid
  sphere in a nonuniform flow, \textit{Phys. Fluids}, \textit{26}, 883.

\bibitem[{\textit{Maximenko et~al.}(2012)\textit{Maximenko, Hafner, and
  Niiler}}]{Maximenko-etal-12}
Maximenko, A.~N., J.~Hafner, and P.~Niiler (2012), {Pathways of marine debris
  derived from trajectories of Lagrangian drifters}, \textit{Mar. Pollut.
  Bull.}, \textit{65}, 51--62.

\bibitem[{\textit{Niiler and Paduan}(1995)}]{Niiler-Paduan-95}
Niiler, P.~P., and J.~D. Paduan (1995), {Wind-driven Motions in the
  northeastern Pacific as measured by Lagrangian drifters}, \textit{J. Phys.
  Oceanogr.}, \textit{25}, 2819--2830.

\bibitem[{\textit{Provenzale}(1999)}]{Provenzale-99}
Provenzale, A. (1999), Transport by coherent barotropic vortices, \textit{Annu.
  Rev. Fluid Mech.}, \textit{31}, 55--93.

\bibitem[{\textit{Ripa}(1997)}]{Ripa-JPO-97b}
Ripa, P. (1997), ``{I}nertial'' oscillations and the $\beta $-plane
  approximation({s}), \textit{J. Phys. Oceanogr.}, \textit{27}, 633--647.

\bibitem[{\textit{Stommel}(1966)}]{Stommel-66}
Stommel, H. (1966), \textit{The {G}ulf {S}tream}, 2nd ed., University of
  California.

\bibitem[{\textit{Sybrandy and Niiler}(1991)}]{Sybrandy-Niiler-91}
Sybrandy, A.~L., and P.~P. Niiler (1991), {WOCE/TOGA Lagrangian drifter
  contruction manual}, \textit{Tech. Rep. SIO Reference 91/6}, Scripps
  Institution of Oceanography, La Jolla, California.

\bibitem[{\textit{Tanga and Provenzale}(1994)}]{Tanga-Provenzale-94}
Tanga, P., and A.~Provenzale (1994), Dynamics of advected tracers with varying
  buoyancy, \textit{Physica D}, \textit{76}, 202--215.

\bibitem[{\textit{Vallis}(2006)}]{Vallis-06}
Vallis, G.~K. (2006), \textit{Atmospheric and oceanic fluid dynamics},
  Cambridge University.

\bibitem[{\textit{{van Sebille} et~al.}(2012)\textit{{van Sebille}, England,
  and Froyland}}]{vanSebille-etal-12}
{van Sebille}, E., E.~H. England, and G.~Froyland (2012), Origin, dynamics and
  evolution of ocean garbage patches from observed surface drifters,
  \textit{Environ. Res. Lett.}, \textit{7}, 044,040.

\end{thebibliography}

\end{article}

\end{document}